\input{aipcheck}
\documentclass[final,sort&compress,numberedheadings]{aipproc}
\layoutstyle{8x11single}
\usepackage{amssymb}
\usepackage{amsmath}
\usepackage{bm}
\newcommand{\be}{\begin{equation}}
\newcommand{\ee}{\end{equation}}
\newcommand{\ba}{\begin{eqnarray}}
\newcommand{\ea}{\end{eqnarray}}
\renewcommand{\l}{\label}
\newcommand{\f}{\frac}

\renewcommand{\a}{\alpha}
\renewcommand{\b}{\beta}
\renewcommand{\d}{\delta}
\newcommand{\g}{\gamma}
\newcommand{\p}{\partial}
\renewcommand{\le}{\left}
\renewcommand{\r}{\right}
\newcommand{\aj}{{\it Astron. J. (USA)}}
\newcommand{\mnras}{{\it Mon. Not. Roy. Astron. Soc. (London)}}
\newcommand{\ncb}{{{\it Nuovo Cimento} B}}
\newcommand{\pla}{{{\it Phys. Lett.}  A}}
\newcommand{\plb}{{{\it Phys. Lett.}  B}}
\newcommand{\prl}{{\it Phys. Rev. Lett.}}
\newcommand{\prd}{{{\it Phys. Rev.} D}}
\newcommand{\pr}{{\it Phys. Rev.}}
\newcommand{\prep}{{\it Phys. Reports}}
\newcommand{\apj}{{\it Astrophys. J.}}
\newcommand{\apjl}{{\it Astrophys. J. Lett.}}
\newcommand{\sa}{{\it Sov. Astron.}}
\newcommand{\grg}{{\it Gen. Rel. Grav.}}
\newcommand{\cm}{{\it Cel. Mech.}}
\newcommand{\cmda}{{\it Cel. Mech. Dyn. Astron.}}
\newcommand{\jmp}{{\it J. Math. Phys.}}
\newcommand{\aap}{{\it Astron. Astrophys.}}
\newcommand{\cqg}{{\it Class. Quant. Grav.}}
\newcommand{\lrr}{{\it Living Reviews in Relativity}}
\newcommand{\ijmp}{{\it Int. J. Mod. Phys.} D}

\def\be{\begin{equation}}
\def\ee{\end{equation}}
\def\bea{\begin{eqnarray}}
\def\eea{\end{eqnarray}}

\begin{document}
\title{Relativistic Reference Frames for Astrometry and Navigation in the Solar System}
\author{ Sergei M. Kopeikin
\footnote{E-mail: kopeikins@missouri.edu}}
{address={Department of Physics \& Astronomy, University of Missouri-Columbia,
Columbia, Missouri 65211, USA}}

\begin{abstract}
Astrophysical space missions
deliver invaluable information about our universe,
 stellar dynamics of our galaxy, and motion of celestial bodies in the solar system. Astrometric space missions SIM and Gaia will determine distances to stars and cosmological objects
as well as their physical characteristics and positions on the
celestial sphere with microarcsecond precision. These and other space missions dedicated to exploration of the solar system are invaluable for experimental testing of general relativity. 
Permanently growing accuracy of space  and ground-based
astronomical observations require corresponding development of 
relativistic theory of reference frames, motion of celestial bodies, and propagation of light/radio signals from a source of light/radio to
observer. Such theory must be based on Einstein's general
relativity and account for various relativistic effects both in
the solar system and outside of its boundary. We describe a hierarchy of the relativistic frames adopted by the International Astronomical Union in 2000, and outline directions for its theoretical and practical extentions by matching the IAU 2000 reference frames in the solar system to the cosmological Friedman-Robertson-Walker reference frame and to the frames used in the parametrized post-Newtonian formalism.
\end{abstract}
\maketitle

\section{Introduction}
\noindent
The role of inertial reference frames in astronomy has been recognized early by both theorists and observers. Astrometric and navigation data rely on observations done in a local inertial frame and transformed to a global inertial frame. The choice of the local reference frame is usually driven by instrumental considerations while construction of the global frame is based upon deeper theoretical grounds. This paper deals with the latter subject.

The original approaches to construct reference frames in astronomy
were completely based on the concepts of the Newtonian gravity and
Euclidean absolute space and time \cite{kom,ws,kovs}. Modern astrometry and navigation,
however, are operating at the angular resolution already exceeding
1 milliarcsecond \cite{min}, and in some cases the resolution has achieved the level of 10 microarcseconds ($\mu$as) \cite{fk}. At this level the
primary gravitational theory must be the Einstein's general 
relativity  (GR) with a corresponding replacement of the Euclidean
space and time by the four-dimensional Riemannian space-time
manifold endowed with the null-cone structure. In other words, the theoretical basis of modern astrometry and navigation
must be entirely relativistic \cite{kom,ws,kovs,kaplan1,kaplan2}. Recognition of this fact is rapidly
spreading in the astrometric and navigation community especially after the successful
completion of the HIPPARCOS mission \cite{hip}, materialization of the
International Coordinate Reference Frame in the sky based on radio quasars \cite{icrf1,icrf2}, development of
technologically new ideas and projects both in space (GP-B \cite{gpb}, FAME \cite{fame}, DIVA \cite{diva}, SIM \cite{sim}, Gaia \cite{gaia}, VSOP \cite{vsop}) and in ground-based (IERS \cite{iers}, VLBA \cite{vlba}, SKA \cite{ska}) astrometry, the discussion \cite{iau1} and the adoption \cite{iau2}
by the XXIVth General Assembly (GA) of the International Astronomical
Union (IAU) of a new post-Newtonian framework for the reference frames
in the Solar system. New motivations for extention of the IAU framework appeared after original discovery in 1998 \cite{p-disc} and recent, more robust experimental confirmation \cite{p-conf} of the, so-called, Pioneer anomaly observed in the hyperbolic orbital motion of spacecrafts (especially Pioneer 10 and 11) in deep space of the solar system. This anomaly may reflect a hidden influence of the gravitational field of our Milky Way galaxy and/or cosmological Hubble expansion on the local dynamics inside the solar system \cite{krbr}. Furthermore, a microarcsecond astrometry can deliver a unique opportunity to measure the Galaxy and cosmological parameters (and perhaps gravitational waves from the early universe \cite{pyne1,pyne2}) from angular measurments of the position and proper motion of quasars in the sky.  Thus, the global reference frame used for interpretation of astrometric and/or navigation data should properly incorporate the galactic \cite{kmak} and cosmological effects \cite{kopg,kere,sofk}.

In what follows we shall describe in more detail the initial theoretical 
motivations that stimulated the development of modern relativistic framework for the reference frames in the solar system and new ideas driving the development 
of the theory beyond 2000. We also provide the reader with a brief history of the development of the relativistic reference frames (section 2). Main notations are explained in section 3. Then, in section 4, we give a description of 
the IAU 1991 reference frame framework and compare it in section 5 with the 
present-day formulation adopted by the XXIVth GA of the IAU 2000. Matching the IAU 2000 framework with the parametrized post-Newtonian formalsim being extensively used in experimental gravity research, is given in section 6. Finally, we match the reference frame of the expanding cosmological model to the IAU 2000 resoltuions in section 7.

Transition from the Newtonian concepts towards more profound
theoretical relativistic approach for the construction reference
frames in modern astrometry and celestial mechanics can be traced
back to the period 1975-1992 and is associated with discovery of a binary pulsar PSR 1913+16 by Hulse and Taylor \cite{ht} and the breakthrough
in the solution of two major theoretical problems: (1) development of a
self-consistent framework for derivation of the post-Newtonian
equations of translational and rotational motion of
self-gravitating extended bodies in the solar system
\cite{asb,k88,bk89a,bk89}, and (2) development of the higher-order
post-Newtonian celestial mechanics of binary pulsars
\cite{d,s,k85}. The main concern of theorists working on relativistic
celestial mechanics of the solar system bodies was related to the
problem of unambiguous interpretation of astronomical measurements
and separation of small coordinate (gauge) perturbations from the real
physical relativistic effects. Later on, the question regarding the
best choice of the gauge conditions imposed on the metric tensor components
arose, for a number of gauges were used in calculations by
various groups, so that there were some arguments about their advantages and
disadvantages as well as about the uniqueness of the results \cite{br79}. But, probably the most serious was the problem of
construction of the local geocentric (and planetocentric)
reference frame in the post-Newtonian approximation(s). Apparently
there existed principal difficulties in solving these problems.
This was because the simplest post-Newtonian approximation of the Earth's
gravitational field by the Schwarzschild spherically-symmetric solution could not be
considered as accurate enough due to the noticeable perturbations
to the metric tensor from the Earth's rotation and oblateness as well as from the tidal forces produced by the Moon, Sun, and other planets. It
was also recognized \cite{k88} that the well-known procedure of
construction of the Fermi normal coordinates in the vicinity of a geodesic world-line of a test particle \cite{fnc1,fnc2} implemented by Fukushima \cite{fuk}, and Ashby and Bertotti \cite{asb} for construction of the geocentric coordinate system can not be
applied due to: (1) the ambiguity in choosing the background space-time
manifold, and (2) deviation of the Earth's center-of-mass world-line
from the geodesic motion because of coupling Earth's internal multipole moments with tidal gravitational field of external bodies. 

Lack of a firm theoretical approach to the problem of the construction of a local inertial frame in the neighborhood of a massive, extended celestial body
led theorists (see, e.g., \cite{will,bai,br72,spr,cap,rhw} and references therein) to use a simple (Euclidean-like) spatial translation of the origin of the barycentric reference frame to the Earth's geocenter
in order to make a geocentric reference frame at the
post-Newtonain approximation. However, though such a procedure is
allowed in general relativity because of the coordinate freedom, the Euclidean
translation to the geocenter produces large relativistic effects
having pure coordinate origin that are unobservable \cite{srs}. In
addition, in such frame the geometric shape of the Earth moving
along the elliptic orbit undergoes the Lorentz and gravitational
contractions which must be compensated by spurious internal
stresses in Earth's interior in order to prevent the appearance of unphysical
deformations of its crust. Similar problems were also met in developing general
relativistic celestial mechanics of binary pulsars \cite{k85,d300}.

Recent technological developments in drag-free satellites, clocks, lasers, optical and radio interferometry \cite{leh,hlnt} make it necessary to elaborate the Will-Nordtvedt parametrized post-Newtonian formalism \cite{will} in order to incorporate relativistic effects in local frames constructed around self-gravitating and extended bodies \cite{ssa,klis1,kv}.    
The domain of applicability of the IAU relativistic theory of 
reference frames \cite{iau2} should be also extended outside the boundaries of the solar system \cite{kopg}.
 
New generation of astrometric satellites which include  SIM \cite{sim},
and Gaia \cite{gaia}(recently adopted as a cornerstone mission of ESA) requires an absolutely new approach for unambiguous
interpretation of astrometric data obtained from the on-board
optical interferometers. SIM and Gaia complement one another -- SIM excels at studying bright and faint sources to high accuracy, while Gaia will obtain astrometric measurements of a large number of stars. SIM represents an entirely new measurement capability in astronomy and a revolution in astrometry that will exceed all previous measurements by a factor of 100-1,000. For faint stars, SIM's capability will exceed that of upcoming missions such as Gaia by a factor of 25-75 or more, and will aproach to 1 $\mu$as. At this level of accuracy
the problem of propagation of light rays must be solved with taking into account relativistic effects generated by non-stationarity of the gravitational field of the solar and other astronomical systems emitting gravitational waves \cite{kopg}. That will allow us to
integrate the equations of light propagation in gravitational field of light-ray-deflecting  bodies without artificial assumptions about position and velocity of the bodies from which it deflects light. Astrometric resolution in 1 $\mu$as
changes the treatment of the parallax, aberration, and proper motion
of observed celestial objects which becomes much more involved and requires better
theoretical definition of reference frames on curved space-time manifold \cite{ks,km,kkp}. This problem is highly releveant to unambigious physical interpretation of the gravitomagnetic precession of GP-B gyroscopes \cite{gpb} which is measured relative to a binary radio star IM Pegasi \cite{bartel} that is pretty close to the solar system (has parallax) and moves rather fast (has large proper motion).

The other motivation for improvement of the theory of reference frames within framework of general relativity is associated with continuously growing precision of the Global Positioning Systsem (GPS) that is a satellite-based navigation system made up of a network of 24 satellites placed into orbit by the U.S. Department of Defense and orbiting the earth about 12,000 miles above its surface. Ephemeris data tells the GPS receiver where each GPS satellite should be at any time throughout the day. Each satellite transmits ephemeris data showing the orbital information for that satellite and for every other satellite in the system. The GPS uses accurate, stable atomic clocks in satellites and on the ground to provide world-wide position and time determination. The best available current precision of the system in range measurements is about 1 cm, and it may be improved by a factor of 10 with making use of radio links of multiple frequencies. This precision suggests that the GPS data processing must include a number of relativistic effects \cite{gps} and definition of geodetic parameters must rely upon general theory of relativity \cite{kman}.

Additional motivation for improving  relativistic  theory of reference 
frames can be relegated to the problem of calculation and detection of the incoming gravitational-wave signals of the space interferometric gravitational wave detectors 
like LISA \cite{lisa} which would consist of three satellites flying apart in space separated 
by distances of order $~1.5\times 10^6$ km in an equilateral triangle. The center of LISA's triangle will follow Earth's orbit around the Sun, trailing 20 degrees behind. The spacecraft will carry instruments to track each other and measure passing gravitational waves. Each of these instruments is made up of two optical assemblies, which contain the main optics, lasers, and a free-falling gravitational reference sensor. LISA will detect gravitational wave sources from all directions in the sky. These sources will include all the thousands of compact binary systems containing neutron stars, black holes, or white dwarfs in our own Galaxy, plus merging supermassive black holes in distant galaxies. However, the detection and proper interpretation of the gravitational-wave signals can be done only under the condition that all coordinate-dependent (or gauge-dependent) effects are completely understood and subtracted from the signal. This is especially important for observation of gravitational waves from very distant sources located at cosmological distances because the Hubble expansion of our universe will affect the propagation of the waves. One should also not forget that LISA would fly in 
the near-zone of the Sun which is as a source of low-frequency gravitational 
waves produced by oscillations of its interior (known as g-modes \cite{curt}). Motion of the LISA satellites and propagation of light rays along the baseline of LISA in the field of such oscillating g-modes must be carefully studied in order to provide a correct interpretation of observations as well.

Many other relativistic experiments for testing general relativity and alternative 
theories of gravity also demand advanced theory of relativistic reference 
frames (see, for example, \cite{kv,ciuf1,ciuf2,llr} and references therein). Two space laser interferometric missions -- ASTROD \cite{astr} and LATOR \cite{latr} -- are dedicated to measure general relativity in higher-order, non-linear regime. 

\section{Brief Historical Remarks}
\noindent
Historically it was de Sitter \cite{desit1,desit2} who worked out a relativistic approach to build reference frames in general relativity. He succeeded in the derivation of relativistic equations of motion of the solar system bodies and discovered the main post-Newtonian effects including the gravitomagnetic perihelion precession of planetary orbits due to the angular 
momentum of the Sun (spin-orbital momentum coupling). He also discovered the geodetic (de Sitter-Fokker) precession of the lunar orbit due to the orbital-orbital momentum coupling that has been verified with the precision of $\leq 1\%$ \cite{berc,sh,tur}. These effects are 
essential in the present-day definition of the dynamic and/or kinematic rotation of a reference frame \cite{bk89,iau2}. 

Later on, Lense and Thirring \cite{lt} gave a more general treatment of the gravitomagnetic dragging of a satellite orbiting around a massive rotating body but they hardly believed that the effect can be measured in practice. It was Ginzburg \cite{g} who first realized that the Lense-Thirring precession can indeed be measured using the earth-orbiting artificial satellites and paid particular attention to the problem of separation of coordinate and physical effects. But only recently the experimental verification of the Lense-Thirring effect has come about with a geodetic satellite LAGEOS \cite{ciuf1,ciuf2,asnat}. In 1959, Pugh \cite{pugh} proposed testing general relativity by observing the precession of a gyroscope with respect to a distant star in a drag-free satellite. Schiff \cite{sff} made calculations and came to the conclusion that a gyroscope in polar orbit at 400 miles should turn with the Earth through an angle amounting after one year to 42 milliarc-seconds due to the Lense-Thirring precession. An interesting generalization of Schiff's calulations was provided by Teyssandier \cite{tey} who worked out a realistic geophysical model that enables us to evaluate the contribution of the earth's nonsphericity to the Lense-Thirring precession of a gyroscope. The gravitomagnetic frame-drag effect is being currently tested by Everitt's GP-P team \cite{gpb}. Terrestrial-based measurment of the gravitomagnetic precession in a laboratory at the south pole station was proposed by Braginskii et al. \cite{brag} and this experiment may be feasible with the current-day technologies. Frame dragging effect caused by translational motion of a massive body has been recently measured in VLBI experiment \cite{fk} that observed a minute relativistic correction to the static-field Shapiro time delay, caused by a moving gravitational field of Jupiter \cite{kapjl,kijmp}.

Ginzburg's paper \cite{g} and the success of the soviet space program motivated Brumberg to develop a post-Newtonian Hill-Brown theory of motion of the Moon \cite{b,bi1} and major planets \cite{bi2} in the solar barycentric coordinate system (see also \cite{br72,br91}). Baierlein \cite{bai} extended  Brumberg's approach accounting for the eccentricity of Earth's orbit. A different approach to the problem of motion of the Moon and the construction of reference frames in general relativity was suggested by Mashhoon and Theiss in a series of papers (for a review see \cite{mat,mt} and references therein). Instead of making use of the standard post-Newtonian approximations (PNA) they developed a post-Schwarzschild treatment of gravitomagnetic effects in the three-body problem. This approach treats gravitational field of the Earth in terms of the exact Schwarzschild solution on which small tidal perturbations from external bodies (Sun, Moon, etc.) are superimposed. It allowed them to discover that the validity of PNA is restricted in time due to the appearance of time-divergent integrals, and the geodetic and gravitomagnetic precessions are parts of the more general phenomena involving a long-term relativistic nutation known as the Mashhoon-Theiss anomaly \cite{gsof}. The same authors also introduced a precise definition of the local geocentric frame with the Earth considered as a massive monopole particle moving on a background space-time manifold \cite{mt,mash}, thus, extending the concept of the Fermi normal coordinates usually introduced in the neighborhood of a massless observer \cite{mtw}. 

A more general approach to the problem of construction of reference frames in general relativity was initiated in our papers \cite{k88} (see also \cite{k87}), where the matched asymptotic expansion technique and a multipolar decomposition of gravitational fields of bodies were employed. This approach has been further refined and used for the development of the, so-called, Brumberg-Kopeikin (BK) formalism for building relativistic astronomical frames in the solar system (or in any other N-body system). This formalism relies upon Einstein's equations, which are solved for the construction of local geocentric and global 
barycentric reference systems whose properties are determined by the specific form of the metric tensor. The metric-tensor matching technique is used for setting up relativistic transformations between these two systems \cite{bk89,bk89a,br91} which generalize the Lorentz transformation of special relativity to the case of presence of gravitational field. 

Damour, Soffel and Xu (DSX) extended the Brumberg-Kopeikin 
theory by protracting the Newtonian definitions of the multipole moments of gravitating bodies used by Brumberg and Kopeikin to the post-Newtonian approximation of general relativity \cite{dsx1,dsx2,dsx3}. Combined BK-DSX formalism made impact on the formulation of resolutions of the General Assembly of the IAU 1991 on reference frames \cite{iau1,a4}. Further lively discussion on the reference frames among members of a joint committee on relativity \cite{jcr} of the Bureau International des Poids et Mesures (BIPM) and of the International Astronomical Union made the complete BK-DSX theory accepted by the XXIV-th General Assembly of the IAU 2000 \cite{iau2} as a basic framework for setting up relativistic time scales and astronomical reference frames in the solar system. 

Further development of the reference frame theory goes to the extention of the BK-DSX theory to alternative theories of gravity \cite{klis1,kv} and matching this theory with the cosmological solution which accounts for the Hubble expansion \cite{kere,kram,ramk}. The BK-DSX theory has been supplemented with the precise mathematical solution of the problem of propagation of test particles and light rays in multipolar gravitational fields of moving bodies \cite{ksge,ks,km,kkp,kjmp,bkk,kquad,klko,klaj,lint,lpl,cm} that allows us to apply that theory for the practical aims of astrometry and navigation in the solar system. 

Recognizing these goals and taking into account that relativitistic questions played an important role in the work of several IAU Commissions and IAU Working Groups for an extended period of time, the XXVI-th General Assembly of the IAU, held in Prague in Aufust 2006, has decided to establish a new IAU Commission 52 "Relativity in Fundamental Astronomy" (RIFA). The general scientific goals of the new commission are
\begin{itemize}
\item clarify geometric and dynamical concepts of fundamental astronomy within relativistic framework;
\item provide adequate mathematical and physical formulations to be used in Fundamental Astronomy;
\item deepen understandings of the above results among astronomers and students in astronomy;
\item promote research needed to acomplish these tasks.
\end{itemize}
This paper pays specific attention to all these issues.

\section{Notations}\l{not}

Latin indices takes values 1,2,3, and the Greek ones run from 0 to 3. Repeated indices indicate the Einstein summation rule. The Kroneker symbol (unit matrix) is denoted $\delta_{ij}={\rm diag}(1,1,1)$ and the fully anti-symmetric symbol of Levi-Civita $\epsilon_{ijk}$ is defined in such a way that $\epsilon_{123}=1$. The Minkowski metric is $\eta_{\a\b}={\rm diag}(-1,1,1,1)$. Greek indices are rised and lowere with the Minkowski metric. By convention, Latin indices are raised and lowered with the Kroneker symbol that makes no difference between upper and lower Latin indices.

Bold italic letters (both upper- and lower-case) denote spatial vectors, for instance ${\bm x}=x^i=(x^1,x^2,x^3)$. Dot and cross between two spatial vectors denote the Euclidean scalar and vector products respectively: ${\bm A}\cdot{\bm B}=A_iB^i=A_iB_i=A^iB^i$, and $({\bm A}\times{\bm B})^i=\epsilon_{ijk}A^jB^k$. 

Partial derivative with respect to spatial coordinates $x^i$ are denoted as $\p/\p x^i$ or ${\vec\nabla}$. Partial derivative with respect to four-dimentional coordinates $x^\a$ are denoted with comma: $F_{,\a}\equiv \p F/\p x^\a$. Total derivative of function, let say, $F(t,{\vec x})$ with respect to time is denoted with overdot, that is $\dot F=dF/dt$, and its partial derivative with respect to time is $\p F/\p t$. Notice that $dF/dt\not=\p F/\p t$. 

$G$ is the universal gravitational constant, and $c$ is the fundamental speed of the Minkowski space-time.

\section{The IAU 1991 Reference Systems Framework}
\noindent
Official transition of the astronomical community from the Newtonian
to relativistic concepts commenced in 1991 when a few 
recommendations (resolution A4) were adopted by the GA of the IAU \cite{iau1,a4}.

In the first recommendation, the metric tensor in space-time
coordinates $(ct,{\bm x})$ centered at the barycenter of an
ensemble of masses was recommended to be written in the form
\begin{eqnarray}
g_{00} &=& -1
+ \frac{2U(t,{\bm x})}{c^{2}} + O(c^{-4})\;,\label{1}\\
\label{2}
g_{0i} &=& O(c^{-3})\;,\\
\label{3}
g_{ij} &=& \delta_{ij}\,\left[1+\frac{2U(t,{\bm x})}{c^{2}}\right] +O(c^{-4})\;,
\end{eqnarray}\noindent
where $c$ is the fundamental speed of the Minkowski space-time (= speed of light in vacuum),
$U$ is the sum of the gravitational potentials of
the ensemble of masses and of a tidal potential generated by bodies
external to the ensemble, the latter  vanishing at the
barycenter. This recommendation recognizes that space-time cannot be described by
a single coordinate system. The recommended form of the metric tensor
can be used not only to describe the barycentric celestial
reference system (BCRS) of the whole
solar system, but also to define the geocentric
celestial reference system (GCRS)
centered on the center of mass of the Earth
with a suitable function $\hat{U}$, now depending upon geocentric coordinates.
In
analogy to the GCRS a corresponding celestial
reference (planetocentric) system can be constructed
for any other body of the Solar system.

In the second recommendation, the origin and orientation of the
spatial coordinate grids for the BCRS and GCRS were defined.
The third recommendation defined $t\equiv TCB$ (Barycentric Coordinate Time) and
$T\equiv TCG$ (Geocentric Coordinate Time) -- the time coordinates of the BCRS and
GCRS, respectively. The relationship
between $TCB$ and $TCG$ is given by a full 4-dimensional transformation\noindent
\begin{equation}\label{4}
TCB - TCG= c^{-2}\,
\left[\int^{t}_{t_0}
\left(\f{v_{E}^2}{2}
      + \bar{U}(t,{\bm x}_E) \right)  dt
+ v^i_E r^i_E \right] + O(c^{-4}),
\end{equation}\noindent
where $x^i_{E}\equiv {\bm x}_E(t)$ and $v^i_{E}$ are the barycentric coordinate position
and velocity of the geocenter, $r^i_E = x^i - x^i_{E}$ with ${\bm x}$ being the
barycentric position of the observer, and $\bar{U}(t,{\bm x}_E)$
is the Newtonian potential of all solar system bodies evaluated at the geocenter apart from that of the Earth. Fundamental and practical issues associated with the new resolutions on the relativistic reference frames and time scales in the solar system were featured in \cite{bkts,sb,sfuk,fuki}.

\section{The IAU 2000 Reference Systems Framework}
\noindent
In August 2000, new resolutions on
reference frames and time scales in the solar system were
adopted by the XXIVth General Assembly of the IAU \cite{iau2}. The resolutions are based on the first post-Newtonian
approximation of general relativity theory and completely abandon the Newtonian paradigm of
space and time that changes terminology in fundamantal astronomy \cite{iau06}. In this section we reproduce the main results of the BK-DSX theory. For more details, the reader is referred to \cite{iau2} and the original BK-DSX publications.

\subsection{Conventions for the Barycentric Celestial Reference System}
\noindent  
Barycentric Celestial Reference System (BCRS) is denoted $x^\a=(ct,{\bm x})$.
BCRS is defined mathematically in terms of a metric tensor $g_{\a\b}$ which components read
\begin{eqnarray}
g_{00} &=& - 1 + \f{2 w}{c^2} - \f{2w^2}{c^4} + O(c^{-5})\;,
\label{5} \\
g_{0i} &=& - \f{4 w^i}{c^3} + O(c^{-5})\;,\label{6}
 \\
g_{ij} &=& \delta_{ij}
\left( 1 + \f{2w}{c^2} \right) + O(c^{-4})\; .
\label{7}
\end{eqnarray}\noindent
Here, the post-Newtonian gravitational potential $w$ generalizes the
usual Newtonian potential $U$ and $w^i$ is the vector potential related
with gravitomagnetic effects.

This form of the barycentric metric tensor implies that the barycentric
spatial coordinates $x^i$ satisfy the harmonic gauge condition \cite{mtw}.
The main arguments in favor of the harmonic gauge are: (1)
tremendous amount of work on isolated astronomical systems in general relativity has been done with the harmonic
gauge that was found to be a useful and simplifying gauge for various kinds
of applications, and (2)
in contrast to the gauge adopted in the PPN formlism and advocated, for example, by Will in his textbook \cite{will}, the harmonic gauge can be defined to higher PN-orders by making use of the rather straighforward, post-Newtonian approximations \cite{d300}, and, in fact, for the exact Einstein theory of gravity.

Assuming space-time to be asymptotically flat (no gravitational fields exist
at infinity) in the standard harmonic gauge the post-Newtonian
field equations of general relativity read
\begin{eqnarray}
\label{eq1}
\Box w&=&-4\pi G\sigma\;,\\
\label{eq2}
\Box w^i&=&-4\pi G\sigma^i\;,
\end{eqnarray}
where $\Box\equiv -c^{-2}\partial^2/\partial t^2+\nabla^2$ is the (gravity-field) wave operator
$\sigma = c^{-2}(T^{00} + T^{ss}),$ $
\sigma^i = c^{-1}T^{0i}$, and
$T^{\mu\nu}$ are the components of the stress-energy tensor of the solar system bodies, $T^{ss}= T^{11} + T^{22} + T^{33}$.

Equations (\ref{eq1}), (\ref{eq2})  
are solved by the post-Newtonian iterations, and in the barycentric coordinate system $(t, x^i)$ one has
\begin{eqnarray}
w(t,{\bm x}) &=& G \int  \;
\f{\sigma(t, {\bm x}')d^3 x'}{\vert {\bm x} - {\bm x}' \vert}
 + \f{G }{2c^2}  \f{\partial^2}{\partial t^2}
\int d^3 x' \, \sigma(t,{\bm x}') \vert {\bm x} - {\bm x}' \vert +O\left(c^{-4}\right)\; ,
\label{8}\\\label{9}
w^i(t,{\bm x}) &=& G \int\; \f{\sigma^i (t,{\bm x}') d^3 x'}{
\vert{\bm x} - {\bm x}' \vert } +O\left(c^{-2}\right)\;,
\end{eqnarray}\noindent
which are to be substituted to the metric tensor (\ref{5})--(\ref{7}). Each of the potentials $w$ and $w^i$ can be linearly decomposed in two parts
\ba\l{eq3}
w&=&w_E+{\bar w}\;,\\\l{eq4}
w^i&=&w_E^i+{\bar w}^i\;,
\ea
where $w_E$ and $w_E^i$ are geopotentials depending on the distribution of Earth's mass and current density expressed in the BCRS, and  ${\bar w}_E$ and ${\bar w}_E^i$ are gravitational potentials of all external bodies but the Earth.

\subsection{Conventions for the Geocentric Celestial Reference System}
\noindent
Geocentric Celestial Reference System (GCRS) is denoted $X^\a=(cT, {\bm X})$.
GCRS is defined in terms of the geocentric metric tensor $G_{\a\b}$ with components represented in the following form
\begin{eqnarray}
G_{00} &=& - 1 + \f{2 W}{c^2} - \f{2W^2 }{ c^4} + O(c^{-5})\;,
\label{11} \\\label{12}
G_{0i} &=& - \f{4 W^i }{c^3} + O(c^{-5})\;,
 \\
G_{ij} &=& \delta_{ij}
\left( 1 + \f{2 W }{ c^2} \right) + O(c^{-4})\; .
\label{13}
\end{eqnarray} \noindent
Here $W = W(T,{\bm X})$ is the post-Newtonian gravitational potential in
the geocentric system, and $W^i(T,{\bm X})$ is the corresponding vector
potential. They satisfy to the same type of the wave equations (\ref{eq1}), (\ref{eq2}) as the corresponding potentials $w$, and $w^i$ in BCRS.
 
These geocentric potentials should be split into two parts: potentials
$W_E$ and $W_E^i$ arising from the gravitational action of the Earth
and external parts ${\bar w}$ and ${\bar w}^i$ due to dynamic (tidal gravity force) and
kinematic (inertial motion) effects. The external parts are assumed to vanish at the
geocenter and admit an expansion into positive powers of ${\bm X}$.
Explicitly,
\begin{eqnarray}
W(T,{\bm X}) &=& W_E(T,{\bm X})
+ W_{\rm kin}(T,{\bm X})+W_{\rm dyn}(T,{\bm X})
\;,
\label{14}\\
\label{15}
W^i(T,{\bm X}) &=& W^i_E(T,{\bm X})
+ W^i_{\rm kin}(T,{\bm X})+W^i_{\rm dyn}(T,{\bm X})
\;.
\end{eqnarray}\noindent
The Earth's geopotentials  $W_E$ and $W^i_E$
are defined in the same way as $w_E$ and $w_E^i$ (see equations (\ref{8})--(\ref{9})) but with quantities $\sigma$ and $\sigma^i$
calculated in the GCRS. $W_{\rm kin}$ and $W_{\rm kin}^i$ are kinematic contributions that are linear in spatial coordinates ${\bm X}$
\begin{equation}\label{16}
W_{\rm kin}= Q_i X^i,
\qquad
W^i_{\rm kin} = \f14\;c^2
\varepsilon_{ipq} (\Omega^p - \Omega^p_{\rm prec})\;X^q\; ,
\end{equation}\noindent where
$Q_i$ characterizes the minute deviation of the actual worldline of
the origin of the GCRS from geodesic motion in the
external gravitational field of the solar system bodies 
\begin{equation}\label{17}
Q_i=\partial_i {\bar w}({\bm x}_E)-a_E^i+O(c^{-2})\;.
\end{equation}\noindent
Here
$a_E^i={dv^i_E/dt}$ is the
barycentric acceleration of the
origin of the GCRS (geocenter). The function
$\Omega^a_{\rm prec}$ describes the relativistic precession
of dynamically nonrotating spatial axes of GCRS with respect to remote celestial objects (quasars)
\begin{eqnarray}\label{18}
\Omega_{\rm prec}^i =
\f{1}{c^2}\,
\varepsilon_{ijk}\,
\left(
-\f32\,v^j_E\,\partial_k {\bar w}({\bm x}_E)
+2\,\partial_k {\bar w}^j({\bm x}_E)
-\f12\,v^j_E\,Q^k
\right).
\end{eqnarray}\noindent
The three terms on the right-hand side of this equation
represent the geodetic, Lense-Thirring, and Thomas precessions,
respectively.  One can prove that $\Omega^a_{\rm iner}$ is dominated by
geodetic precession  amounting to $\sim 2^{\prime\prime}$ per century
plus short-periodic terms usually called geodetic nutation.
One sees that for $\Omega^i = \Omega^i_{\rm prec}$ the kinematic vector potential
$W^i_{\rm kin}$ vanishes. This implies that dynamical equations of
motion of a test body, e.g., a satellite orbiting around the Earth, do not contain the
Coriolis and centrifugal terms, i.e., the local geocentric spatial
coordinates ${\bm X}$ are {\it dynamically non-rotating}. For practical reasons,
however, the use of {\it kinematically non-rotating} geocentric
coordinate axes defined by condition $\Omega^i = 0$ is recommended by IAU. {\it Kinematically non-rotating} GCRS keeps its spatial axes fixed with respect to the axes of the BCRS while {\it dynamically non-rotating} GCRS makes spatial axes slowly precessing with respect to the BCRS.

Potentials $W_{\rm dyn}$ and $W^i_{\rm dyn}$ are generalizations
of the Newtonian tidal potential. 
For example, 
\begin{equation}\label{19}
W_{\rm dyn}(T,{\bm X}) =
{\bar w}({\bm x}_E + {\bm X}) - {\bar w}({\bm x}_E) - X^i \p_i{\bar w}({\bm x}_E)+O(c^{-2}).
\end{equation}
It is easy to check out that a Taylor expansion of ${\bar w}({\bm x}_E + {\bm X})$ around the point ${\bm x}_E$ gives a polynom starting from the quadratic with respect to ${\bm X}$ terms.
We also note that the local gravitational potentials $W_E$ and $W_E^i$ of the
Earth are related to the barycentric gravitational potentials $w_E$ and
$w^i_E$ by the relativistic transformations \cite{iau2}
\begin{eqnarray}\label{20}
W_E(T,{\bm X})&=&w_E(t,{\bm x})\,\left(1+\f{2v_E^2}{ c^2}\right)-
\f{4}{ c^2}\,v_E^i\,w^i_E(t,{\bm x})+O(c^{-4}),
\\\label{21}
W^i_E(T,{\bm X})&=&R^i{}_j\,\left[w^j_E(t,{\bm x})
-v_E^j\,w_E(t,{\bm x})\right]+O(c^{-2}).
\end{eqnarray}\noindent
where $R^i{}_j(t)$ is a (time-dependent) orthogonal matrix of rotation (with the angular velocity $\Omega^i_{\rm prec}$) arising because of the choice of the {\it kinematically non-rotating} GCRS. In case of the {\it dynamically  non-rotating} GCRS the matrix $R^i{}_j(t)=\delta^i_j$.

\subsection{Transformations between the Reference Systems}
\noindent
The coordinate transformations between the BCRS and GCRS are found by matching the BCRS and GCRS metric tensors in the vicinity of the world line of the Earth by making use of their tensor properties. The transformations are written as
\begin{eqnarray}
\label{22}
T&=&t - \f{1}{ c^2} \left[ A(t) + {\bm v}_E\cdot{\bm r}_E \right]
+ \f{1}{ c^4} \left[ B(t) + B^i(t)\,r_E^i +
B^{ij}(t)\,r_E^i\,r_E^j + C(t,{\bm x}) \right] +O(c^{-5}),
\\\label{23}
X^i&=&
r^i_E+\frac 1{c^2}
\left[\frac 12 v_E^i {\bm v}_E\cdot{\bm r}_E + {\bar w}({\bm x}_E) r^i_E
+ r_E^i {\bm a}_E\cdot{\bm r}_E-\frac 12 a_E^i r_E^2
\right]+O(c^{-4}),
\end{eqnarray}\noindent
where ${\bm r}_E={\bm x}-{\bm x}_E$, and functions $A(t), B(t), B^i(t), B^{ij}(t), C(t,{\bm x})$ are
\def\xe{({\bm x}_E)}
\begin{eqnarray}
\f{dA(t)}{dt}&=&\f12\,v_E^2+{\bar w}\xe,\label{24}
\\
\f{dB(t)}{ dt}&=&-\f{1}{8}\,v_E^4-\f{3}{ 2}\,v_E^2\,{\bar w}\xe
+4\,v_E^i\,{\bar w}^i+\f{1}{2}\,{\bar w}^2\xe,\label{25}
\\\label{26}
B^i(t)&=&-\f{1}{2}\,v_E^2\,v_E^i+4\,{\bar w}^i\xe-3\,v_E^i\,{\bar w}\xe,
\\\label{27}
B^{ij}(t) &=& -v_E^{i} Q_{j}+
2 \p_j {\bar w}^i({\bm x}_E)
-v_E^{i} \p_j {\bar w}\xe+\f{1}{ 2}
\,\delta^{ij} \dot{{\bar w}}({\bm x}_E),
\\\label{28}
C(t,{\bm x})&=&-\f{1}{10}\,r_E^2\,({\dot{\bm a}_E}\cdot{\bm r}_E)\, .
\end{eqnarray}\noindent
Here again $x_E^i, v_E^i$, and $a_E^i$ are the barycentric position,
velocity and acceleration vectors of the Earth, the dot stands for the
total
derivative with respect to $t$. 
Let us also remark that the harmonic gauge
condition does not fix the function $C(t,{\bm x})$ uniquely. However, it is reasonable to fix it in the time transformation for practical reasons.

\section{Matching the IAU 2000 Framework with the PPN Formalism}
\noindent
General theory of relativity is the most powerful theoretical tool for experimental gravitational physics both
in the solar
system and outside of its boundaries. It passed all tests with
unparallel degree of accuracy \cite{willrv,schaefer,dam2000}.
However, alternative theoretical models are required for deeper
understanding of the nature of space-time gravitational physics and for studying possible
violations of general relativistic relationships which may be observed
in
near-future gravitational experiments.

The goal of this section is to discuss how to incorporate the parametrized post-Newtonian (PPN) formalism \cite{will} to the
IAU 2000 theory of general relativistic reference frames in the solar
system \cite{iau2}.  This will extend domain of applicability of the resolutions
to more general class of gravity theories. Furthermore, it will make
the IAU 2000 resolutions fully compatible with the JPL equations of motion used for calculation of ephemerides of major
planets, Sun and Moon.
These equations depend on two PPN parameters, $\beta$ and $\gamma$ \cite{sman} and they
are presently compatible with the IAU 2000 resolutions only in the case of $\beta=\gamma=1$.
Rapidly growing precision of optical and radio astronomical
observations as well as calculation of relativistic equations of motion in gravitational wave astronomy urgently
demands to work out a PPN theory of relativistic
transformations between the local and global coordinate systems.

PPN parameters $\beta$ and $\gamma$ are characteristics of a hypothetical scalar field which perturbs the metric tensor and makes
it different from general relativity.
Scalar fields have not yet been detected but they already play
significant role in modern physics for many reasons \cite{sfi1,sfi2}. In order to extend the IAU 2000 theory of reference frames to the PPN formalism we employed a general class of the scalar-tensor theories of gravity initiated in the pioneering works by
Jordan \cite{jo1,jo2}, Fierz \cite{frz} and, especially, Brans
and Dicke \cite{brd,dicke,dicke1}.
This class of theories is
based on the metric tensor $g_{\alpha\beta}$ representing gravitational field and a scalar
field $\phi$ that couples with the metric tensor through the coupling
function $\theta(\phi)$ which we keep arbitrary. Following Will \cite{will}, we assume that $\phi$ and $\theta(\phi)$ are analytic functions
which can be expanded about their cosmological background values $\bar{\phi}$ and $\bar{\theta}$. Existence of
the scalar field $\phi$
brings about dependence of the universal gravitational constant $G$ on
the background value of the field $\bar{\phi}$ which can be
considered as constant on the time scale much shorter than the Hubble
cosmological time.

\subsection{Local Coordinate System in the PPN Formalism} 

Our purpose is to develop a theory of relativistic reference
frames in an N-body problem (solar system) with two parameters $\beta$ and
$\gamma$ of the PPN formalism. There is a principal difficulty in developing such a theory associated with
the problem of construction of a local coordinate system in the vicinity of each
self-gravitating body (Sun, Earth, planet) comprising the N-body system.
Standard textbook on the PPN formalism \cite{will} does not contain solution of this problem because the
original PPN formalism was constructed in a single, asymptotically flat,
global PPN coordinates covering the entire space-time and having the
origin at the barycenter of the solar system. PPN formalism admits existence of several fields which are responsible
for gravity -- scalar, vector, tensor, etc. After imposing boundary conditions on all these fields at infinity the
standard PPN metric tensor merges their contributions all together in a single expression so that they get absorbed
to the Newtonian and other general relativistic potentials and their contributions are strongly
mixed up. Therefore, it is technically impossible in standard PPN formalsim to disentangle the different fields in order to find out relativistic space-time
transformation between the local
frame of a self-gravitating body (Earth) and the global PPN coordinates which would be consistent with the law of
transformation of the fields imposed by each specific theory of gravitation.

It is quite straightforward to construct the post-Newtonian Fermi coordinates along a world line of a massless particle
\cite{fnc1}. Such approach can be directly applied in the PPN formalism to construct the Fermi reference
frame around a world line of, for example, an artificial satellite. However, account for gravitational self-field
of the particle (extended body) changes physics of the problem and introduces new mathematical aspects to the
existing procedure of construction of the Fermi frames as well as to the PPN formalism.
Only three papers \cite{ssa,klis1,kv} have been published so far where
possible
approaches aimed to derive the
relativistic transformations between the local (geocentric,
planetocentric) and the PPN global coordinates were discussed in the framework of the PPN formalism.
The approach proposed in \cite{ssa}
is based on the formalism that was originally worked out by Ashby and
Bertotti \cite{asb} and Fukushima \cite{fuk} in order to construct a local
inertial frame in the vicinity of a self-gravitating body that is
a member of an N-body system. In the Ashby-Bertotti formalism
the PPN metric
tensor is taken in its standard form \cite{will} and it treats all massive bodies
as point-like monopole massive particles without rotation. Construction of
a local inertial frame in the vicinity of such massive particle
requires to impose some specific restrictions on the world line of the particle. Namely,
the particle is assumed to be moving along a geodesic defined on the ``effective"
space-time manifold which is obtained by elimination of the body under
consideration from the expression of the standard PPN metric tensor. This
way of introduction of the ``effective" manifold is not
defined uniquely
bringing about an ambiguity in the construction of the ``effective"
manifold that was noticed by Kopeikin \cite{k88}. Moreover, the assumption that bodies
are point-like and non-rotating is insufficient for modern geodesy and relativistic celestial mechanics \cite{kman}.
For example,
planets in the solar system and stars in binary systems have appreciable rotational speeds and noticeable higher-order
multipole moments. Gravitational interaction of the multipole moments of a celestial body with external tidal field does
not allow the body to
move along the geodesic line \cite{k88}. Deviation of the body's center-of-mass world line from the geodesic
motion can be significant and important in numerical calculations of planetary ephemerides \cite{kman,standish}) and must be taken into account when one constructs a theory of the
relativistic reference frames in the N-body system.

Different approach to the problem of construction of a local
(geocentric) reference frame in the PPN formalism was proposed in
the paper by Klioner and Soffel \cite{klis1}. These authors have
used a phenomenological approach which
does not assume that the PPN metric tensor in local coordinates must be a solution of the field equations of a
specific
theory of gravity. The intention was to make this kind of formalism as general as possible. To this end, these
authors assumed that the structure of the metric tensor written down in the local
(geocentric) reference frame must have the following properties:
\begin{itemize}
\item[{\bf A.}]
gravitational field of external with respect to the Earth bodies (Sun, Moon,
planets) is represented in the vicinity of the Earth in the form
of tidal potentials which should reduce in the Newtonian limit to the Newtonian tidal potential,
\item[{\bf B.}]
switching off the tidal potentials must reduce the metric
tensor of the local coordinate system to its standard PPN
form.
\end{itemize}
Direct calculations revealed that under assumptions made in \cite{klis1} the
properties (A) and (B) can not be satisfied simultaneously. This
is a direct consequence of abandoning the field equations for construction of the metric tensor in geocentric coordinates and the matching procedure applied in \cite{klis1} in order
to transform the local geocentric
coordinates to the global barycentric ones. More specifically, at each step of
the matching procedure of \cite{klis1} four kinds of different terms in the metric tensors have been
singling out and equating separately in the corresponding matching equations for the
metric tensor:
\begin{itemize}
\item the terms depending on internal potentials of the body under consideration (Earth);
\item the terms which are functions of time only;
\item the terms which are linear functions of the local spatial
coordinates;
\item the terms which are quadratic and higher-order polynomials of the
local coordinates.
\end{itemize}
These matching conditions are implemented
in order to solve the matching equations. 

We draw attention of the reader to the problem of choosing the right number of the matching equations. In general
theory of relativity the only gravitational field variable is the metric tensor. Therefore, it is necessary and
sufficient to write down the matching equations for the metric tensor only. However, alternative theories of gravity
have additional fields (scalar, vector, tensor) which contribute to the gravitational field as well. Hence, in these
theories one has to use matching equations not only for the metric tensor but also for these fields. The paper \cite{klis1} assumes that it is sufficient to solve the matching
equations for the metric tensor only in order to obtain complete information about the structure of the
parametrized post-Newtonian transformation from the local to global frames but this is not correct as shown in \cite{kv}. In our approach \cite{kv} we consistently use the
matching equations for both
the metric tensor and the scalar field which are derived from the field equations.
Klioner - Soffel approach to the
PPN formalism with local frames taken into account admits too many degrees of freedom in matching equations which can not be fixed uniquely.  Restriction of this freedom
can be done {\it ad liberum}, thus leading to researcher-dependent ambiguity in the construction and physical interpretation of relativistic effects in the local (geocentric) reference frame.

Our point of view is that in order to eliminate any ambiguities in the construction
of the PPN
metric tensor in the local reference frame of the body under consideration and to apply mathematically rigorous
procedure for derivation of the
relativistic coordinate transformations from the local to global
coordinates, a specific theory of gravity must be used. This fixes the structure of the field equations and the number of functions entering the PPN metric tensor in the local coordinates that becomes exactly equal to the number
of matching equations. Hence, all of them can be determined uniquely. We propose to build a parametrized
theory of relativistic reference
frames in the solar system by making use of the following procedure:
\begin{enumerate}
\item chose a class of gravitational theories with a well-defined field equations;
\item impose a specific gauge condition on the metric tensor and other fields to single out a set of global and
local coordinate systems and to reduce the field equations to a solvable form;
\item solve the reduced field equations in the global (barycentric) coordinate system $x^\alpha=(ct,x^i)$ by imposing fall-off boundary conditions at infinity;
\item solve the reduced field equations in the local (geocentric) coordinate system $X^\alpha=(cT,X^i)$ defined in the vicinity of the Earth;
\item make use of the residual gauge freedom to eliminate non-physical degrees of freedom and to find out the most
general structure of the space-time coordinate transformation between the barycentric and geocentric coordinates;
\item transform the metric tensor and the other fields from the local coordinates to the global ones by making use
of the general form of the coordinate transformations found at the previous step;
\item derive from this transformation a set of matching (the first-order differential and/or algebraic) equations for
all functions entering the metric tensor and the coordinate transformations;
\item solve the matching equations and determine all functions entering the matching equations explicitly.
\end{enumerate}\noindent
This procedure works perfectly in the case of general relativity \cite{iau2} and is valid in the class of the
scalar-tensor theories of gravity as well \cite{kv}. 

The scalar-tensor
theory of gravity employed in \cite{kv} operates with one tensor,
$g_{\alpha\beta}$, and one scalar, $\phi$, fields. The tensor
field $g_{\alpha\beta}$ is the metric tensor of the Riemannian
space-time manifold. The scalar field $\phi$ is
generated by matter of the bodies comprising an N-body system. We assume that the N-body system consists of extended bodies which moves slowly and have weak
gravitational field. We use the post-Newtonian
approximation (PNA) scheme 
in order to
find solutions of the scalar-tensor field equations. It takes into account the
post-Newtonian definition of gravitational multipole moments introduced by Thorne \cite{thorne} and
further elaborated by Blanchet and Damour \cite{bld}. We do not specify the internal structure of the bodies so that our
consideration is not restricted with the case of a perfect fluid that extends the PPN formalism \cite{will}. 

\subsection{Transformations between the Reference Systems in the PPN Formalism}
\noindent
The coordinate transformations between the BCRS and GCRS are found by matching the BCRS and GCRS metric tensors and the scalar field in the vicinity of the world line of the Earth. The transformations have the following form \cite{kv}
\begin{eqnarray}
\label{22q}
T&=&t - \f{1}{ c^2} \left[ A(t) + {\bm v}_E\cdot{\bm r}_E \right]
+ \f{1}{ c^4} \left[ B(t) + B^i(t)\,r_E^i +
B^{ij}(t)\,r_E^i\,r_E^j + C(t,{\bm x}) \right] +O(c^{-5}),
\\\label{23q}
X^i&=&
r^i_E+\frac 1{c^2}
\left[\frac 12 v_E^i {\bm v}_E\cdot{\bm r}_E + \gamma{\bar w}({\bm x}_E) r^i_E
+ r_E^i {\bm a}_E\cdot{\bm r}_E-\frac 12 a_E^i r_E^2
\right]+O(c^{-4}),
\end{eqnarray}\noindent
where ${\bm r}_E={\bm x}-{\bm x}_E$, and functions $A(t), B(t), B^i(t), B^{ij}(t), C(t,{\bm x})$ are
\def\xe{({\bm x}_E)}
\begin{eqnarray}
\f{dA(t)}{ dt}&=&\f{1}{2}\,v_E^2+{\bar w}\xe,\label{24q}
\\
\f{dB(t)}{ dt}&=&-\f{1}{8}\,v_E^4-\le(\gamma+\f{1}{ 2}\r)\,v_E^2\,{\bar w}\xe
+2(1+\gamma)\,v_E^i\,{\bar w}^i+\le(\beta-\f{1}{ 2}\r)\,{{\bar w}}^2\xe,\label{25q}
\\\label{26q}
B^i(t)&=&-\f{1}{2}\,v_E^2\,v_E^i+2(1+\gamma)\,{\bar w}^i\xe-(1+2\gamma)\,v_E^i\,{\bar w}\xe,
\\\label{27q}
B^{ij}(t) &=& -v_E^{i} Q_{j}+
(1+\gamma) \p_j {\bar w}^i({\bm x}_E)
-\gamma v_E^{i} \p_j {\bar w}\xe+\f{1}{ 2}
\,\delta^{ij} \dot{{\bar w}}({\bm x}_E),
\\\label{28q}
C(t,{\bm x})&=&-\f{1}{10}\,r_E^2\,({\dot{\bm a}_E}\cdot{\bm r}_E)\, .
\end{eqnarray}\noindent
Here again $x_E^i, v_E^i$, and $a_E^i$ are the barycentric position,
velocity and acceleration vectors of the Earth, the dot stands for the
total
derivative with respect to $t$. 

As one can see the PPN transformations between the geocentric and barycentric coordinates depend explcitly on two PPN parameters $\beta$ and $\gamma$. The gauge function $C(t,{\bm x})$ remains
unaffected by these parameters. In case of general relativity, $\beta=\gamma=1$, and the PPN transformations (\ref{22q})--(\ref{28q}) are reduced to the form adopted by the International Astronomical Union and given in equations (\ref{22})--(\ref{28}).

The PPN geocentric coordinates do not rotate {\it kinematically} with respect to the International Celestial Reference System (ICRF) defined by a set of primary reference quasars \cite{icrf1,icrf2}. {\it Dynamically non-rotating} geocentric coordinates are precessing with respect to ICRF with the angular velocity \cite{kv}
\begin{eqnarray}\label{18q}
\Omega_{\rm prec}^i =
\f{1}{c^2}\,
\varepsilon_{ijk}\,
\left[
-\le(\gamma+\f12\r)\,v^j_E\,\partial_k {\bar w}({\bm x}_E)
+(\gamma+1)\,\partial_k {\bar w}^j({\bm x}_E)
-\f12\,v^j_E\,Q^k
\right],
\end{eqnarray}\noindent
which depends on the PPN parameter $\gamma$. Equation (\ref{18q}) is similar to that derived in \cite{mtw} for a test spinning particle (gyroscope) moving in external gravitational field of a rotating mass. We emphasize that equation (\ref{18q}) has been derived for a massive self-gravitating body as opposed to the case of the test particle. Equation (\ref{18q}) goes to general relativistic expression (\ref{18}) when $\gamma=1$.

\section{Matching the IAU 2000 Framework with Cosmological Reference Frame}
\noindent
The rapidly growing accuracy of astronomical measurements makes it necessary to take into account some important cosmological effects for an adequate interpretation of optical and radio observations of cosmological lenses, anisotropy in cosmic microwave background radiation, etc. Some, yet unexplained anomalies in the orbital motion of the solar system bodies \cite{p-conf,krbr,dittus} may be associated with the cosmological expansion and/or other cosmology-related effects (multidimensional branes, strings). For this reason, matching of the IAU 2000 framework for reference frames in the solar system with the cosmological coordinates of the Friedman-Robertson-Walker (FRW) universe \cite{mtw} becomes vitally important. In this section we outline the main ideas for doing this matching that incorporates the Hubble expansion effects to the post-Newtonian approximation of general relativity worked out previously for an asymptotically flat space-time. They were worked out in our papers \cite{kram,ramk}.

The main observation is that our universe is not asymptotically-flat and, hence, gravitational field does not vanish as one approaches infinity. The assumption of vanishing gravitational field and its first derivatives was extensively used in the Newtonian and relativistic celestial mechanics of N-body system but, in fact, it is not realistic and should be replaced with the cosmological boundary conditions. To this end, we considered a FRW universe driven by quintessence (that is, a scalar field imitating the dark matter \cite{sfi2}) $\phi$, and having a spatial curvature equal to zero \cite{sfi1,sfi2}. The background universe is perturbed by a localized, self-gravitating system of N bodies (the solar system). In this model the perturbed metric tensor reads 
\begin{equation}
\label{c1}
g_{\alpha\beta}=a^2(\tau)(\eta_{\alpha\beta}+h_{\alpha\beta})\;,
\ee
where perturbation $h_{\alpha\beta}$ of the background FRW metric tensor $\bar g_{\a\b}=a^2\eta_{\a\b}$ is caused by the presence of the solar system, $a(\tau)$ is a scale factor of the universe depending on cosmological time $\tau$ related to coordinate time $t$ by simple differential equation $dt=a(\tau)d\tau$. In what follows, a linear combination of the metric perturbations
\be\l{c1a}
\gamma^{\alpha\beta}=h^{\alpha\beta}-\f12\eta^{\alpha\beta}h\;,
\ee
where $h=\eta^{\alpha\beta}h_{\alpha\beta}$, is more convenient for calculations.

We discovered a new cosmological gauge \cite{kram} which has a number of remarkable properties \cite{kopetr}. In case of the background FRW universe with dust equation of state (that is, the background pressure of matter is zero \cite{sfi1,sfi2}) this gauge is given by \cite{kram,ramk} 
\begin{equation}
\label{c2}
\gamma^{\a\b}{}_{|\b}=2H\varphi \d^\a_0\;,
\end{equation}
where bar denotes a covariant derivative with respect to the background metric $\bar g_{\a\b}$, $\varphi=\phi/a^2$, $H=\dot a/a$ is the Hubble parameter, and the overdot denotes a time derivative with respect to time $\tau$. The gauge (\ref{c2}) generalizes the harmonic gauge of asymptotically-flat space-time for the case of the expanding non-flat background universe. The most attractive property of the new gauge is that it allows us to decouple the system of the Einstein equations for the perturbations of gravitational field directly in the position space without making their Fourier transform in a series of "plane waves" as it is usually done in standard cosmological perturbation theory \cite{sfi2}.

The gauge (\ref{c2}) drastically simplifies the linearized Einstein equations which are reduced to the following form
\begin{eqnarray}\label{c3}
\Box\gamma_{\alpha\beta}-2H\;\partial_\tau\gamma_{\alpha\beta}+H^2\le[\d^0_\a\g_{b0}+\d^0_\b\g_{\a0}+\d^0_\a\d^0_\b(\g-2\varphi)\r]=-\f{16\pi G}{c^4} T_{\alpha\beta}\;,
\end{eqnarray}
where $\partial_\tau\equiv\p/\p\tau$,  $\,\Box\equiv -c^{-2}\p^2_\tau+{\bm\nabla}^2$, $\gamma\equiv -\gamma_{00}+\gamma_{kk}$, and $T^{\alpha\beta}$ is the tensor of energy-momentum of matter of the solar system. Introducing new notations $\gamma_{00}\equiv 4w/c^2$, $\gamma_{0i}\equiv -4w^i/c^3$, and $\gamma_{ij}\equiv 4w^{ij}/c^4$, and spliting Einstein's equations in time-time, time-space, and space-space components, yield
\begin{eqnarray}\label{c5}
\Box\chi-2H\p_\tau \chi+\f52H^2\chi&=&-4\pi G\sigma\;,
\\\label{c6}
\Box w-2H\p_\tau w\phantom{oooo[[::::}&=&-4\pi G\sigma-4H^2\chi\;,\\
\label{c7}
\Box w^{i}-2H\p_\tau w^{i}+H^2 w^{i}&=&-4\pi G\sigma^{i}\;,\\
\label{c8}
\Box w^{ij}-2H\p_\tau w^{ij}\phantom{ooooo}&=&-4\pi G T^{ij}\;,
\end{eqnarray}
where $\chi\equiv w-\varphi/2$, the Hubble parameter $H=\dot a/a=2/\tau$, and densities $\sigma=c^{-2}(T^{00}+T^{ss})$ and $\sigma^i=c^{-1}T^{0i}$. These Bessel-wave equations extend the domain of applicability of the post-Newtonian equations (\ref{eq1}), (\ref{eq2}) to the case of expanding universe. 

Equations (\ref{c5})--(\ref{c8}) consist of eleven equations as besides the metric tensor we have a scalar field $\phi$ that mimics a dark matter. First equation (\ref{c5}) describes evolution of the scalar field while the second equation (\ref{c6}) describes evolution of the scalar perturbation of the metric tensor. Equation (\ref{c7}) yileds evolution of vector perturbations of the metric tensor, and equation (\ref{c8}) describes generation and propagation of gravitational waves by the isolated N-body system. Equations (\ref{c5})--(\ref{c8}) 
contain all corrections depending on the Hubble parameter and can be solved analytically in terms of generalized retarded solution. Exact Green functions for these equations have been found in \cite{kram,ramk} (see also \cite{pois1}). They revealed that the gravitational perturbations of the isolated system on expanding background depend not only on the value of the source taken on the past null cone but also on the value of the gravitational field inside the past null cone. Existence of extra terms in the solutions of equations  (\ref{c5})--(\ref{c8}) depending on the Hubble parameter brings about cosmological corrections to the Newtonian law of gravity. For example, the post-Newtonian solutions of equations (\ref{c6}), (\ref{c7}) with a linear correction due to the Hubble expansion are  
\ba\l{cor1}
w(\tau,{\bm x}) &=& G \int  \;
\f{\sigma(\tau, {\bm x}')d^3 x'}{\vert {\bm x} - {\bm x}' \vert}
 + \f{G }{2c^2}  \f{\partial^2}{\partial \tau^2}
\int d^3 x' \, \sigma(\tau,{\bm x}') \vert {\bm x} - {\bm x}' \vert - GH\int d^3 x'\sigma(\tau, {\bm x}') +O\left(c^{-4}\right)+O\left(H^2\right)\; ,
\\\label{cor2}
w^i(\tau,{\bm x}) &=& G \int\; \f{\sigma^i (\tau,{\bm x}') d^3 x'}{
\vert{\bm x} - {\bm x}' \vert } +O\left(c^{-2}\right)+O\left(H^2\right)\;.
\end{eqnarray}\noindent
Linear with respect to the Hubble parameter, $H$, term is absent in equation (\ref{cor2}) due to the law of conservation of linear momentum of an isolated self-gravitating system. 
Matching these solutions with those defined in the BCRS of the IAU 2000 framework in equations (\ref{8}), (\ref{9}) is achieved after expanding all quantities depending on the conformal time $\tau$ in the neighbourhood of the present epoch in powers of the Hubble parameter.

\begin{theacknowledgments}
This work is completed with support of the Eppley Foundation for Research, Inc. (New York) (project \#C0006269). I thank J. Kovalevsky, T. Fukushima, G. Petit, and P. Teyssandier for valuable comments. 
\end{theacknowledgments}
\bibliographystyle{aipprocl}

\end{document}